\documentstyle[a4wide]{article}

\begin{document}
\makebox[14cm][r]{TTP 94-13}\par
\makebox[14cm][r]{hep-ph/9410260}\par
\makebox[14cm][r]{LNF-94/066(IR)}\par
\makebox[14cm][r]{May 1995}\par
\vspace{.7cm}
\centerline{\Large Tau Decays into Four Pions}
\par
\vspace{1.cm}
\centerline{R. Decker, P. Heiliger and H. H. Jonsson}
\par \centerline{Institut f\"ur Theoretische Teilchenphysik} \par
\centerline{Universit\"at Karlsruhe}\par
\centerline{D--76128 Karlsruhe, Germany}\par \par
\vspace{0.5cm}
\centerline{M. Finkemeier\footnote{%
Address from July 1995: Lyman Laboratory of Physics,
Harvard University, Cambridge MA 02138, USA}} \par
\centerline{INFN - Laboratori Nazionali di Frascati} \par
\centerline{P.O. Box 13} \par
\centerline{I--00044 Frascati (Roma), Italy} \par
\normalsize
\vspace{2.cm}

\begin{abstract}
We compute branching ratios and invariant mass distributions of the
tau decays into four pions. The hadronic matrix elements are
obtained by starting from the structure of the hadronic current in
chiral limit and then implementing low-lying resonances in the
different channels.
Reasonable
agreement with experiment is obtained both for the $\tau \to
\nu_{\tau} + (4 \pi \!) $ decay rates and the $e^+e^- \to (4 \pi \!)
$  cross sections.
Furthermore we supply an interface to use
our matrix elements within the Tauola Monte-Carlo program.
\end{abstract}

\newpage
\noindent
\section{Introduction}
The last two years have provided us with a vast of new data on the
physics of the $\tau $ lepton. (See \cite{helt} for a review of the
most recent experimental results.) The overall branching ratio
deficit that has been a problem for many years is getting smaller
and considerable improvement in the detection of small
exclusive branching ratios involving $\pi^0 $'s and kaons has been
achieved. In general agreement between theory and experiment is
satisfactory.

In this paper we concentrate on a specific hadronic final state,
namely the $\tau $ decays into four pions, $\tau^- \to \nu_{\tau}
\pi^- \pi^- \pi^+ \pi^0 $ and $\tau^- \to \nu_{\tau} \pi^0 \pi^0
\pi^0 \pi^- $. These decays are related through CVC to
corresponding $e^+ e^- $ data. The predictions \cite{tsai,gil,eidel} from
$e^+ e^- $ data for the total branching ratio are in good agreement
with the $\tau $--data \cite{helt,arg,cleo,aleph}.

However,
this ansatz allows only to calculate the integrated rates.
In this paper we will develop a dynamical model
which allows us to describe differential decay rates.

In our discussion we follow along the lines set in the
description for $\tau $ decays into three
pseudoscalars \cite{rog}. Lorentz--invariance dictates the number of
relevant form factors. Then the resonances dominating these form
factors are identified (in our case namely the $\rho$, $\rho'$,
$\rho''$, $a_1$ and $\omega $ mesons) and an expression for the
form factors is constructed. Furthermore the expressions are
restricted by assuming that only on mass-shell vertices
contribute and that different vertices are related by Breit--Wigner
propagators. In the case of the $\tau$ decays into three
pseudoscalars, the coupling constants at the vertices were assumed
to be constant, and the products of the coupling constants were
fixed by the requirement that the form factors have the correct
chiral limit \cite{wess}.
In the case of the four pion final states, we will be
forced to assume that these products are slowly varying functions of
the momentum, because otherwise we cannot reproduce the experimental
data.

An expression constructed from the chiral limit and including $\rho
$ resonances in all possible ways had been constructed and included
in TAUOLA \cite{tauo}. Unfortunately it yields a branching ratio off
by about a factor of 2 to 4. In this paper we improve this result by
including {the $3 \pi $ resonance, namely the $a_1 $ meson},
and obtain reasonable agreement with experiment.

In section 2 we discuss the general structure of the $\tau $ decay
into four pions and we implement the low--lying resonances according
to the different $4 \pi $ final states into the hadronic matrix
element. We compare the model predictions with a model--independent
determination of the relative probabilities of the channels
corresponding to the alternatives of charged and neutral $\pi $'s
\cite{pais}. In section 3 we discuss the two possible $4 \pi $ final
states in $e^+ e^- $ annihilation and the relationship of these
processes to the $\tau $ decays into four pions. We discuss the
numerical results as well as the invariant mass--distributions in
the various $2 \pi $--subsystems in comparison to experiment in
section 4.

\section{The Four Pion Decay Mode of the Tau}
There exist two $4 \pi $ final states in $\tau^- $ decay:
\begin{itemize}
\item[i)] $\tau^- (P) \to \nu (l) \pi^0 (p_1) \pi^0 (p_2) \pi^0 (p_3)
           \pi^- (p_4) $
\item[ii)] $\tau^- (P) \to \nu (l) \pi^- (p_1) \pi^- (p_2) \pi^0
           (p_3) \pi^+ (p_4) $
\end{itemize}
In the standard model \cite{sm} the following effective lagrangian is
responsible for the semileptonic decays of the $\tau $ lepton
\cite{rog}
\begin{equation}
L_{eff} = {G_F U^{ij} \over \sqrt{2}} {\overline u}(l') \gamma_{\mu }
(1 - \gamma_5) u(l) H_{ij}^{\mu }
\end{equation}
with $G_F $ the Fermi constant, $H_{ij}^{\mu } $ and $U^{ij} $ are
the charged quark current and the quark mixing matrix respectively.
In the following we present a model for the matrix element of the
quark current, which reveals itself as being in agreement with
general properties of a system of multi--pions. These considerations
are consequences of a classification of various pion--subsystems
according to isospin \cite{pais}. In our model the charged current
consists of three parts which we explain in detail below.

Our model is based on an amplitude derived \cite{wess} in the chiral
limit, supplemented by vector dominance to correctly describe the
resonances in the $4 \pi $ mass distributions and a slow $Q^2$
dependence of the meson couplings. In this frame
% Change in 3rd version
we calculate $\tau \to 4 \pi \nu_{\tau} $ via sequential processes like
$\tau \to \nu_{\tau} \rho, \; \rho \to a_1 \pi, \; a_1 \to \rho \pi,
\; \rho \to \pi \pi $ and $\tau \to \nu_{\tau} \rho, \; \rho \to
(\pi \pi)_{s-wave} \rho, \; \rho \to \pi \pi $. To construct the
covariant amplitudes corresponding to Fig. 1 we need the appropriate
vertices and propagators. Following the successful reasoning in a
similar analysis concerning tau decay into three pseudoscalar mesons
\cite{rog} we assume the vertex functions to be given by their on
shell structure and to be transverse, e.g.
\begin{equation}
p_{a_1}^{\nu} \Gamma_{\nu \sigma} (a_1 \to \rho \pi ) =
p_{\rho}^{\sigma} \Gamma_{\nu \sigma} (a_1 \to \rho \pi) =
p_{\rho}^{\tau} \Gamma_{\tau} (\rho \to \pi \pi) = 0
\end{equation}
With the pions all on mass shell, the vertex factors may thus be
written \cite{isg}
\begin{eqnarray}
\label{eqn3}
\Gamma^{\mu \nu} \left (a_1(q) \to \rho(k) \pi(p) \right ) & = &
f_{a_1 \rho \pi} (q^2,k^2) \left ( - g^{\mu \nu} + {k^{\mu} k^{\nu}
\over k^2} + {q^{\mu} q^{\nu} \over q^2} - {k \cdot q \over k^2 q^2}
q^{\mu} k^{\nu} \right ) \nonumber \\
\Gamma^{\nu} \left ( \rho (k) \to \pi_1 (p_1) \pi_2 (p_2) \right )
& = & -i f_{\rho \pi \pi} (k^2) (p_1^{\nu} - p_2^{\nu })
\end{eqnarray}
where the
meson couplings are real.
In \cite{rog} these meson couplings were assumed to be constant.
In the current case of $\tau\to 4\pi \nu_\tau$, however, where the momentum
transfers involved can be relatively high, we will be forced to weaken
this assumption and allow for a slow variation with the momentum
transfers, see below.
In the first case, we have neglected a further possible
coupling which is of higher order in momenta.

The resonances are in general described by normalized Breit--Wigner
resonance propagators $BW_R(Q^2=s) $ with $BW(0) = 1 $ and
\begin{equation}
BW_R(s) = {-M_R^2 \over s - M_R^2 + i \sqrt{s} \Gamma_R(s) }
\end{equation}
where the energy dependent width $\Gamma_R(s) $ is computed from
its usual definition
\begin{equation}
\Gamma_R(s) = {1 \over 2 \sqrt{s}} |M_{R \to f_i}|^2 d\Phi
\delta(Q - \sum p_i) \;\;\;\; \Gamma_R(s)_{s = M_R^2} = \Gamma_R
\end{equation}
with $\Gamma_R(s) = 0 $ for $s < $ threshold. For $R $ decaying into
two particles with masses $M_1, M_2 $, this formula is simplified to
\begin{eqnarray}
\Gamma_R(s) & = & \Gamma_R {M_R^2 \over s} \left ( {p \over p_R}
\right )^{2n+1} \nonumber \\
p & = & {1\over 2 \sqrt{s}} \sqrt{(s - (M_1 + M_2)^2)(s - (M_1 -
M_2)^2)} \nonumber \\
p_R & = & {1\over 2 M_R} \sqrt{(M_R^2 - (M_1 + M_2)^2)(M_R^2 -
(M_1 - M_2)^2)}
\end{eqnarray}
where $n $ is the power of $|p| $ in the matrix element. In the
Breit--Wigner of the $a_1 $ we use
\begin{eqnarray}
m_{a_1} = 1.251\; \mbox{GeV}, \;& &  \Gamma_{a_1} = 0.599 \;
\mbox{GeV} \nonumber
\\
\sqrt{s}\, \Gamma_{a_1} (s) & = & m_{a_1} \Gamma_{a_1} {g(s) \over
g(m_{a_1}^2)}
\end{eqnarray}
where the function $g(s) $ is given by \cite{kuhn}
\begin{equation}
g(s) = \left ( \begin{array}{cc} 4.1(s - 9 m_{\pi}^2)^3(1 - 3.3(s -
9 m_{\pi}^2) + 5.8 (s - 9 m_{\pi^2}^2)^2) & \mbox{if} \; s <
(m_{\rho} + m_{\pi})^2 \\
s(1.623 + {10.38 \over s} - {9.32 \over s^2} + {0.65 \over s^3})
& \mbox{else} \end{array} \right )
\end{equation}
The vector resonances for more meson channels allow a richer
structure, containing the $\rho $--meson and some excitations.
With energy dependent widths we take
\begin{equation}
T^{(i)}_{\rho} (s) = \left \{ {\sum_R \beta_R^{(i)} BW_R^{(i)}(s)
\over \sum_R \beta_R^{(i)}
} \right \}
\end{equation}
The index $i$ indicates the fact that for rho like resonances
which couple to different particles, the relative contributions
$\beta_R^{(i)}$ of the
different radial excitations need not be the same.
For $a_1 \to \rho \pi$ we use the parameterization which was derived
in
\cite{kuhn} and also used in \cite{tauo,rog}:
\begin{eqnarray}
m_\rho = 0.773\; \mbox{GeV}, \; & & \Gamma_\rho = 0.145\;
\mbox{GeV}\nonumber
\\
m_{\rho'} = 1.370\; \mbox{GeV}, \; & & \Gamma_{\rho'} = 0.510 \;
\mbox{GeV}\nonumber
\\
\beta_{\rho'}^{a_1\to\rho\pi} = -0.145
\end{eqnarray}
For all other rho-like resonances we take for the moment a common
$T_\rho$. We fix its parameters (masses, widths and $\beta_R$'s) by
fitting the cross section for the process $e^+ e^- \to 2\pi^+ 2\pi^-
$ to experimental data (see Sec.~4).

\bigskip
These effective vertices and propagators allow to construct the full
amplitude for the $\tau $ decays into $4 \pi $ final states using the
structure of the chiral currents \cite{wess}, and assuming a
dominant role of the $a_1 $ meson in these decays, as indicated by
the data. Explicitly we have for this part of the hadronic current
in eq.(1) the following matrix element:
\begin{eqnarray}
<4 \pi|H^{\mu (1)}|0> & \sim & T_{\rho }(Q^2)\, \Gamma^{\mu \nu}
\left (\rho (Q) \to a_1(k) \pi \right ) \, BW_{a_1}(k^2) \nonumber \\
& & \cdot \Gamma^{\nu \lambda} \, \left (a_1(k) \to \rho(l) \pi
\right ) \, BW_{\rho}(l^2)\, \Gamma^{\lambda} (\rho(l) \to \pi \pi)
\end{eqnarray}
In the chiral limit, where the meson masses tend to zero
and then the limit $s \to 0 $ is performed, the hadronic current
should have the form \cite{wess}
\begin{eqnarray} \label{eqn12}
J_{\mu}(\pi^- \pi_1^0 \pi_2^0 \pi_3^0) & = & {2\sqrt{3} \over
f_{\pi}^2} \sum_{i=1}^3 (q_{0i} - q_- )^{\nu} A_{\mu \nu}^{i-}
\nonumber \\
J_{\mu}(\pi_1^- \pi_2^- \pi^+ \pi^0) & = & {2\sqrt{3} \over f_{\pi}^2
} [ 2(q_+ - q_0)^{\nu} A_{\mu \nu}^{+0} + (q_{-1} - q_+)^{\nu}
A_{\mu \nu}^{1+} + (q_{-2} - q_+)^{\nu} A_{\mu \nu}^{2 +}] \nonumber
\\
\mbox{with} \;\;\;\;\;\;\;\;\;\;\;\;\;\;\;\; & & \\
A_{\mu \nu}^{ik} & = & g_{\mu \nu} - \sum_{l \neq i,k}
{(Q - 2 q_l)_{\mu} (Q - q_l)_{\nu} \over (Q - q_l)^2 } \nonumber
\end{eqnarray}
Explicit calculation shows that the decay chain $\tau \to \nu_{\tau}
\rho, \rho \to a_1 \pi, a_1 \to \rho \pi, \rho \to \pi \pi $ alone
does not have the correct chiral limit. This can be corrected by
adding additional contributions, see Fig. 1 (b), which can be
interpreted as being induced by two pions in the $s $--wave which
possibly induce $H^{\mu (2)} $ given by
\begin{equation}
<4 \pi | H^{\mu (2)} |0> \; \sim \; T_{\rho}(Q^2) \left ( g^{\mu
\nu} - {Q^{\mu} Q^{\nu} \over Q^2 } \right ) T_{\rho} (k^2)
\Gamma^{\nu} (\rho (k) \to \pi \pi ) (BW_{f_0}((Q-k)^2))
\end{equation}
Thus a second decay chain of the type $\tau \to \nu_{\tau} \rho,
\rho \to (\pi \pi)_s \rho, \rho \to \pi \pi $ is built
up; where the total charge of the two pions is zero,
we introduce a Breit--Wigner for $(\pi \pi)_s $
resonance: $f_0 $.
In its Breit-Wigner we use
\begin{equation}
   m_{f_0} = 1300 \, \mbox{MeV}
   \qquad  \Gamma_{f_0} = 600 \, \mbox{MeV}
\end{equation}
We could assume the meson couplings at the vertices (see Eqn.~\ref{eqn3}) to
be constant over the whole range from zero up to the
resonance region, as was done in \cite{rog}. The relevant products of these
coupling constants and the relative contributions of the two decay
chains could then be fixed from the chiral limit, i.e.\ from their values at
zero momentum transfer.
Note in this context
that we assume a complete vector meson dominance in the following
sense. In the chiral limit, the current consists of a direct $W\to4\pi$
part and of a graph where $W\to 2\pi$ and one of the pions converts into
$3\pi$, (the pion pole terms, see also \cite{wess}). We assume that this chiral
current, as given in Eqn.~\ref{eqn12}, is
the low energy limit of the sum of the two graphs in Fig.~1.
Note also that in our framework, the pion pole terms (for massless pions) are
reproduced by the vector meson dominance graphs in Fig.~1, due to the special
ansatz for the meson couplings in Eqn.~\ref{eqn3}.

In matter of fact, however, we find that it is impossible to obtain a
good fit to the $\tau\to 4 \pi \nu_\tau$ data and the
$e^+ e^- \to 4\pi$ cross sections (see Sec.~3), if we assume the meson
couplings to be constant over the whole range of momentum transfers from $0$
up to the tau mass $m_\tau$.
We have tried various modifications of our model
in order to be able to obtain a good fit with constant couplings.
For
example, we allowed for different resonance parameters for the
different $T_\rho^{(i)}$ involved, modified the $T_\rho^{a_1 \to
\rho\pi}$, or assumed that the dominance by the $a_1$ occurs only
for the decay of the $\rho(770)$, whereas for the $\rho'$ we fixed
the $a_1$ resonance factor equal to unity. However, in all these
cases, the best fit for the $e^+e^-$
cross section is at least by about a factor of
two or three too large. Furthermore, when keeping constant couplings,
the
fits tend to prefer unphysical values for the resonance parameters.
Therefore we are convinced that when going from $Q^2 = 0$ to $Q^2 \approx
(1.5\, \mbox{GeV})^2$, the couplings at the vertices can no longer be
assumed to be constant.
This contrasts with the behaviour at lower $Q^2\, {}_\sim^< \, 1
\,\mbox{GeV}^2$, where reasonable descriptions can be obtained using
effective chiral Lagrangians, i.e.\ assuming constant coupling constants
\cite{russen}.

Thus we are forced
to multiply the hadronic currents with an additional function
$F(Q^2)$, which describes the dominant effect of the
energy dependence of the meson couplings.
 From the low energy theorem, we know that it is unity at zero
momentum squared, $F(0) =1$. However, we have found that $F(Q^2)$
can no longer be unity at $Q^2 \sim (1.5 \, \mbox{GeV})^2$. In order
to be able to reproduce the experimental $e^+ e^- \to 4 \pi$ cross
sections, we have to assume that $F(Q^2)$ is considerably lower than
unity at high $Q^2$. Now the question arises which functional
dependence of $F(Q^2)$ should be assumed. From experimental data
\cite{arg} we know that in about $75 \%$ of the decays $\tau^-\to
\pi^-\pi^-\pi^+ \pi^0\nu_\tau$, the invariant mass of the four pions
falls into the relatively small range of $ Q^2 = (1.15 \cdots
1.55\, \mbox{GeV})^2 $. Therefore we make the approximation that
$F(Q^2)$ is constant in the relevant energy interval:
\begin{equation}
    F(Q^2) = \left\{ \begin{array}{cl}
        1 & \mbox{for\ \ }Q^2 = 0 \\
        \gamma & \mbox{for\ \ } Q^2 \approx (1.15 \cdots 1.55\,
        \mbox{GeV})^2
    \end{array} \right.
\end{equation}
So in fact we multiply the hadronic current by a constant $\gamma$,
but we have to keep in mind that this might be a bad approximation
outside the range indicated above. Furthermore the two decay chains
need not be multiplied by the same function $F(Q^2)$. So we could
multiply the first decay chain with the $a_1 \pi$ intermediate state
by $\gamma_1$ and the second decay chain with the $(\pi \pi)_s \rho$
intermediate state by a different $\gamma_2$. For simplicity,
however, we use
\begin{equation}
   \gamma = \gamma_1 = \gamma_2
\end{equation}
Our approach can also be seen from a slightly different point of view.
In the $1 \cdots 2 \, \mbox{GeV}$ region, the cross sections and decay rates
are
dominated by vector and axial vector meson exchange, such as in Figs.~1 and 2.
Therefore a good description can be obtained by using Breit-Wigner forms for
the resonances with couplings defined at the resonances (on-shell
couplings), as was done in \cite{gil,pham}, for example.
In fact, this is also what the approach of the present paper amounts to.
We fix
the relevant products of the on-shell couplings (and some other parameters)
by using experimental data on $e^+ e^- \to 2 \pi^- 2 \pi^+$ and
on $\tau \to \omega \pi \nu_\tau$ (see below).
We have tried to fix the couplings by assuming them to be constant down to zero
momentum transfer and by imposing the chiral limit.
This, however, did not lead to a reasonable description.
The deviation of the parameter $\gamma$ from unity indicates the size of
the variation of the product of couplings, when the momentum transfer goes
from zero up to the $\rho'$ mass of about $1.5 \, \mbox{GeV}$.

In addition to the two decay chains considered till now,
the decay $\tau^- \to 2
\pi^- \pi^+ \pi^0 $ has contributions from an anomalous part ($\rho
\omega \pi $--coupling) \cite{rog2} in Fig. 2; the matrix element of
the corresponding hadronic current reads \cite{tauo}
\begin{eqnarray}
<4 \pi |H^{\mu\; anom} |0> & = & \left ( {1\over m_{\rho}^2 - Q^2 -
i m_{\rho}\Gamma_{\rho} } + \sigma  {1\over m_{\rho'}^2 - Q^2 - i
m_{\rho'} \Gamma_{\rho'} } \right ) \nonumber \\
& & \cos {\Theta_c} G_{\omega 3 \pi} g_{\rho \omega \pi} F_{\rho}
\left \{ {1\over m_{\omega}^2 - (Q-p_2)^2 - i m_{\omega}
\Gamma_{\omega} } \right. \nonumber \\
& & \left [ p_{1 \mu} \left ( (Q - p_2) \cdot p_3 p_4 \cdot p_2
- (Q - p_2) \cdot p_4 p_3 \cdot p_2 \right ) \right. \nonumber \\
& & + p_{3 \mu} \left ( (Q - p_2) \cdot p_4 p_1 \cdot p_2
- (Q - p_2) \cdot p_1 p_4 \cdot p_2 \right ) \nonumber \\
& & \left. + p_{4 \mu} \left ( (Q - p_2) \cdot p_1 p_3 \cdot p_2
- (Q - p_2) \cdot p_3 p_1 \cdot p_2 \right ) \right ] \nonumber \\
& & \left. + (1 \leftrightarrow 2) \right \}
\end{eqnarray}
As for the parameters $\sigma$, $m_{\rho'}$ and $\Gamma_{\rho'}$, we
stick to the parameterization used in \cite{tauo}.
The product $G_{\omega 3 \pi} g_{\rho \omega \pi}$ was taken to be
\begin{equation}
   G_{\omega 3 \pi} g_{\rho \omega \pi}
   = 1476\, \mbox{GeV}^{-3} 12.924\, \mbox{GeV}^{-1}
\end{equation}
in \cite{tauo},
where the coupling constants have been extracted from experimental
data. Note, however, that both numbers have some uncertainty.
Especially the extraction of $g_{\rho\omega\pi}$ from the
$\omega\to\gamma\pi$ decay rate is not free from theoretical
uncertainties. The rate for the decay $\tau\to\pi\omega\nu_\tau$
obtained with the above product disagrees somewhat with the
experimental number \cite{arg}.
Therefore we write
\begin{equation}
   G_{\omega 3 \pi}\, g_{\rho \omega \pi}
   =  { g_\omega \, \cdot ( 1476 \cdot  12.924)\, \mbox{GeV}^{-4}}
\end{equation}
where $g_\omega$ is a number of order one to be determined below.

\section{Relation between the Tau Decays and the $e^+e^-$
Annihilation Cross Sections}
There are two possible $4 \pi $ final states in $e^+ e^- $
annihilation, $2\pi^- 2 \pi^+ $ and $\pi^+ \pi^- 2 \pi^0 $ (the
$4 \pi^0 $ channel is forbidden by charge--conjugation invariance).
They are accessible by a different $I_3 $ component of the same
$I = 1 $ weak current describing $\tau $ decay. This allows to relate
these  processes to the $\tau $ decays \cite{gil}
\begin{eqnarray} \label{eqnee}
{\Gamma (\tau^- \to \nu_{\tau} \pi^- 3\pi^0) \over \Gamma (\tau^-
\to \nu_{\tau} e^- \overline{\nu}_e )} & = &  {3 \cos^2 \theta_c
\over 2 \pi \alpha^2
m_{\tau}^8} \int_0^{m_{\tau}^2} dQ^2 Q^2 (m_{\tau}^2 - Q^2)^2
(m_{\tau}^2 + 2 Q^2) \nonumber \\
& & \cdot \left [{1\over 2} \sigma_{e^+e^- \to 2\pi^- 2\pi^+} (Q^2)
\right ] \nonumber \\
{\Gamma (\tau^- \to \nu_{\tau} 2\pi^- \pi^+ \pi^0) \over \Gamma
(\tau^- \to \nu_{\tau} e^- \overline{\nu}_e) } & = & {3 \cos^2
\theta_c \over 2 \pi
\alpha^2 m_{\tau}^8} \int_0^{m_{\tau}^2} dQ^2 Q^2
(m_{\tau}^2 - Q^2)^2 (m_{\tau}^2 + 2Q^2) \nonumber \\
& & \cdot \left [{1\over 2} \sigma_{e^+e^- \to 2\pi^- 2\pi^+} (Q^2)
+ \sigma_{e^+ e^- \to \pi^+ \pi^- 2\pi^0}(Q^2) \right ]
\end{eqnarray}

These relations can be inverted { (in the range of $Q^2 $ covered
by the tau )} to
$$
%\begin{eqnarray}
  \sigma_{e^+e^- \to 2\pi^+2\pi^-}(Q^2)  =
  {2 \pi \alpha^2\over 3 \cos^2 \theta_c}
  {m_\tau^2 \over Q^3 (1 - Q^2/m_\tau^2)^2
  (1 + 2 Q^2/m_\tau^2)}  \mbox{\ \hspace*{3cm}\ }
$$
%\nonumber \\
$$
  \cdot
  \left[ {\Gamma_{\pi^- 3\pi^0} \over
   \Gamma_e}
  {1\over N_{\pi^- 3\pi^0} }
  {dN_{\pi^- 3\pi^0}(Q)\over dQ} \right]
$$
$$
  \sigma_{e^+ e^- \to \pi^+ \pi^- 2\pi^0}(Q^2)  =
  {\pi \alpha^2\over 3 \cos^2 \theta_c}
  {m_\tau^2 \over Q^3 (1 - Q^2/m_\tau^2)^2
  (1 + 2 Q^2/m_\tau^2)} \mbox{\ \hspace*{3cm}\ }
%\nonumber \\
$$
\begin{equation}
\cdot
  \left[
   {\Gamma_{\pi^+2\pi^-\pi^0} \over
   \Gamma_e}
  {1\over N_{\pi^+2\pi^-\pi^0} }
  {dN_{\pi^+2\pi^-\pi^0}(Q)\over dQ}
   - {\Gamma_{\pi^+2\pi^-\pi^0} \over
   \Gamma_e}
  {1\over N_{\pi^+2\pi^-\pi^0} }
  {dN_{\pi^+2\pi^-\pi^0}(Q)\over dQ} \right]
\end{equation}
(Here $Q$ denotes the square root of the four
momentum squared, $Q := \sqrt{Q_\mu Q^\mu}$.)

Thus any measurement of, and any model for differential
distributions of the tau decays into four pions has implications for
the electron positron annihilation cross sections. Whereas from a
measurement of the tau decays there are of course only implications
for cross sections up to $Q^2 < m_\tau^2$, within a model for tau
decays one can formally assume a larger mass of the tau and deduce
predictions for the cross sections even at higher $Q^2$.

%%%%%%%%%%%%%%%%%%%%%%%%

Before confronting the model with experiment and making more
detailed predictions, it is worth noting that any model for the
description of a multipion final state has to satisfy quite general
conditions of charge correlations following from isospin
considerations \cite{pais}. They fix the relative probabilities of
the channels corresponding to the various alternatives of charged and
neutral $\pi $'s. In the case of the tau decays into four pions
these considerations amount to the statement that
\begin{equation}
{\Gamma (\tau^- \to 2 \pi^- \pi^+ \pi^0) \over \Gamma (\tau^- \to
3 \pi^0 \pi^- )} \ge {3\over 2}
\end{equation}
which is well satisfied in our model, as will be seen in the next
section.

\section{Numerical Results in Comparison to Experiment}

We still have to determine the following parameters of our model,
$g_\omega$, $\beta_{\rho'}$, $\beta_{\rho''}$,
$m_{\rho'}$, $\Gamma_{\rho'}$ and $\gamma \equiv F(Q^2 \approx
m_{\rho'}^2)$. (The parameters of the $\rho'' \equiv \rho(1700)$,
which turns out to contribute only very little, are kept at
their particle data book values \cite{pdg}.) The $\omega$ does not
contribute to $e^+e^- \to  2\pi^+ 2\pi^-$, so the cross section for
this process is independent of $g_\omega$. We determine the other
parameters by fitting the cross-section obtained with our model to
experimental data, see Fig.~3. Our best fit is
\begin{eqnarray}
   \beta_{\rho'} = 0.08 & \qquad \qquad & \beta_{\rho''} = -0.0075
\nonumber \\
   m_{\rho'} = 1.35\, \mbox{GeV} & & \Gamma_{\rho'} = 0.3\,
\mbox{GeV} \nonumber \\
   m_{\rho''} \equiv 1.70\, \mbox{GeV} & & \Gamma_{\rho''}
\equiv 0.235\, \mbox{GeV}
\nonumber \\
   \gamma = 0.38
\end{eqnarray}
where the $\rho''$ parameters have been stated for completeness.

The result of our fit in Fig.~3 lies somewhat below the data in the
range $\sqrt{Q^2} = 1.9\mbox{--}2.2\, \mbox{GeV}$. This could
possibly be cured by including the $\rho(2150)$ in the fit. As our
main goal is to describe the tau decays, where $Q^2$ can of course
not be larger than $m_\tau^2$, we do not do this. Furthermore one has
to remember that at such high $Q^2$, $F(Q^2)$ is presumably no
longer constant.
There is also some disagreement between our fit and the Nowosibirsk
data \cite{nowo} in the region $\sqrt{Q^2} \approx 1.3\, \mbox{GeV}
$, but as the Nowosibirsk and the new Orsay data \cite{orsay} do not
match well, this might be due to errors in the experimental data.

Next we have fixed $g_\omega$ by requiring that the experimental
results for $\tau \to \nu_{\tau} \omega \pi $ \cite{arg} are
reproduced reasonably (see Tab.~\ref{tabbrs} below). This results in
\begin{equation}
   g_\omega = 1.4
\end{equation}
Having fixed the parameters from $e^+e^- \to 2\pi^+ 2\pi^-$ and from
$\tau\to \nu_\tau \omega \pi $, we obtain predictions for the cross
section for $e^+e^-\to \pi^+\pi^-2\pi^0$ and for the other tau
decays into four pions.

Our prediction for the $e^+e^-\to \pi^+\pi^-2\pi^0$ cross section is
shown in Fig.~4. The experimental data for this state are much poorer
than those for the other mode, and the data from different
experiments do not agree very well. Note that our prediction is
rather low, but still compatible with the data. It is interesting
that our prediction favours the new Orsay data \cite{orsay}, which
is in fact also true for the process $e^+e^- \to 2\pi^+ 2\pi^-$,
see Fig.~3.

Now we will discuss the $\tau$ decays. Our results for the integrated
decay rates are given in Tab.~\ref{tabbrs}, where we compare with the
version 2.4 of Tauola and with experimental data.
The branching ratios have been calculated assuming a tau lifetime of
\begin{equation}
   \tau_\tau = 295.7\, \mbox{fs}
\end{equation}
Note that we use physical masses for the pions, $m_{\pi^+}
\neq m_{\pi^0}$, in the phase space,
and  a tau mass of
\begin{equation}
   m_\tau = 1.7771\, \mbox{GeV}
\end{equation}
This explains why the numbers in Tab.~\ref{tabbrs} for Tauola 2.4 do
not agree exactly with those in Tabs.~4 and 5 of \cite{tauo}.
Whereas the numbers for the $\tau\to\nu_\tau\omega\pi$ mode have
been fitted by adjusting $g_\omega$, the other values are
predictions. They agree reasonably well with the experimental values,
although the prediction for $\mbox{BR}(2\pi^- \pi^+ \pi^0)$ appears
to be somewhat too small.

Our predictions are to be compared to the earlier ones in
\cite{gil,eidel,clegg}. While the various predictions for
$\mbox{BR}(\pi^- 3 \pi^0)$ both in the present paper and in
\cite{gil,eidel,clegg} are all very similar, those for
$\mbox{BR}(2\pi^- \pi^+ \pi^0)$
spread quite a lot. This is of course due to the fact that
$\mbox{BR}(\pi^- 3 \pi^0)$ is determined by
$\sigma(e^+ e^- \to 2 \pi^+ 2 \pi^-)$,
which is known quite well experimentally, whereas
$\mbox{BR}(2\pi^- \pi^+ \pi^0)$
depends strongly on
$\sigma(e^+ e^- \to  \pi^+  \pi^- 2 \pi^0)$,
where there are contradicting experimental results.

In \cite{clegg}, where a detailed comparison of $e^+ e^-$ cross sections and
$\tau$ decay rates has been performed, the
data have been analyzed in terms of partitions as
defined by Pais \cite{pais} rather than by a dynamical model.
 From these results, the authors deduce values of the rates
into definite charge states in $\tau $ decay. For $\Gamma(\pi^{\pm}
3 \pi^0)/\Gamma_e$ they obtain a value of $0.050 \pm 0.002 $,
for $\Gamma(\pi^{\pm}
\pi^+ \pi^- \pi^0)/{\Gamma_e}$ their result is
 $ 0.190 \pm 0.009 $ or $0.223 \pm 0.011 $,
according to two different data sets.

In \cite{eidel}, predictions for various tau branching ratios based on
CVC are given,
based on direct integration over the experimental values of $e^+ e^-$
cross sections. Their result for $\mbox{BR}(2 \pi^- \pi^+ \pi^0) = (4.20
\pm 0.29) \%$ is considerably larger than ours.
This is due to our rather small
prediction for the $e^+e^-\to \pi^+\pi^- 2 \pi^0$ cross section,
which is supported, as we have stated before, by the new Orsay data.
Whereas the integration over experimental results in \cite{eidel} takes
all experimental results at face value, including the rather
contradicting data on
$\sigma(e^+ e^- \to  \pi^+  \pi^- 2 \pi^0)$,
our model makes a definite prediction for this cross section, which
favours certain experiments.

{
%%%%%%%%%%%%%%%%%%%%%%%%%%%%%%%%%%%%%%%%%%5
\begin{table}
\caption{Numerical results for the integrated
decay rates of our model, compared to Tauola 2.4 and to experimental
data}
\label{tabbrs}
$$
\begin{array}{|c|c|c|c|}
\hline\hline
  \mbox{observable} & \mbox{present model} & \mbox{Tauola 2.4}
& \mbox{experiment}
 \ \\
\hline
  \Gamma(\tau^-\to\nu_\tau 2\pi^-\pi^+\pi^0) / \Gamma_e
   & 0.172 & 0.0832 &
\\ \hline
 \mbox{BR}(\tau^-\to\nu_\tau 2\pi^-\pi^+\pi^0)
   & 3.11\% & 1.51 \% & (4.25 \pm 0.28) \% \cite{aleph}
\\ \hline
  \Gamma(\tau^-\to\nu_\tau \pi^- 3\pi^0) / \Gamma_e
  & 0.0540 & 0.0172 &
\\ \hline
  \mbox{BR}(\tau^-\to\nu_\tau \pi^- 3\pi^0)   &
  0.98 \% & 0.31\% &
       1.15 \pm 0.15 \% \cite{helt}
\\ \hline
  \Gamma(\tau^-\to\nu_\tau \pi^- \omega(\pi^-\pi^+\pi^0)) / \Gamma_e
 &
  0.0664 & 0.0339 &
\\ \hline
  \mbox{BR}(\tau^-\to\nu_\tau \pi^- \omega(\pi^-\pi^+\pi^0))   &
   1.20 \% & 0.61 \% & 1.65 \pm 0.36 \% \cite{arg}
\\ \hline
\frac{\displaystyle  \Gamma(\tau^-\to\nu_\tau \pi^- \omega(\pi^-
\pi^+\pi^0))}
{\displaystyle \Gamma( \tau^-\to\nu_\tau 2\pi^-\pi^+\pi^0) }
 & 0.39 & 0.41 & 0.33 \pm 0.05 \cite{arg}
\\ \hline
\end{array}
$$
\end{table}
%%%%%%%%%%%%%%%%%%%%%%%%%%%%%%%%%%%%%%%%%%%%%%
%
Our results for the differential distributions in the decay $\tau\to
\nu_\tau 2\pi^-\pi^+\pi^0$ are given in Figs.~5--7. We compare with
the Argus data \cite{arg} and with the 2.4 version of Tauola.
For the Argus data we have performed a rebinning into larger bins.

Note that we display distributions
\begin{equation}
   \frac{1}{N} \frac{dN}{dm}
\end{equation}
which are  normalized to the total number of events $N$.
It is important to stress that the
predicted total rate by TAUOLA 2.4 is much too low, which is
not visible in these normalized shapes.
We run the
Monte-Carlos with very high statistics. Therefore the statistical
errors of the Monte-Carlo results are very small and not shown in
the figures.

Overall we find a good agreement between our model and the
experimental data.
In the invariant mass of the four pion system (Fig.~5), our
prediction for the peak of the distribution is a bit on the high
side, but we do not consider this as very pronounced. We have tried
to improve the fit by lowering the mass of the $\rho'$. This,
however, inevitably leads to predictions for the $e^+e^-$ cross
sections which badly disagree with the experimental data.

In comparing the experimental three pion invariant mass distribution
in Fig.~6 with our model one should remember the limited
experimental mass resolution. Furthermore, we have chosen
$g_\omega = 1.4$ such that both the ratio $\frac{\displaystyle
\Gamma( \tau^-\to\nu_\tau \pi^- \omega(\pi^-\pi^+\pi^0))}
{\displaystyle \Gamma( \tau^-\to\nu_\tau 2\pi^-\pi^+\pi^0) }$ and the
absolute number for the branching ratio agree reasonably with the
experimental numbers. However, in the line shapes obviously only the
relative contribution of the $\omega$ channel is important, which
comes out a bit on the high side with $g_\omega = 1.4$. A somewhat
lower value of about 1.2 or 1.3 would result in a better fit here.

In the case of the two pion invariant mass distributions (Fig.~7
(a)--(e)), we get good agreement for the charged opposite sign
combination (containing the $\rho^0$)(Fig.~7(a)) and the charged like
sign one (Fig.~7(e)).  In the case of the charged-neutral
combinations, however, we find that we predict more  $\rho^+$ than seen
by ARGUS (Fig.~7(d)).
Note, however, that recent Aleph data \cite{aleph} do support our large
prediction for the contribution of $\rho^+$.

\section{Conclusions}
In this paper we have analyzed the four pion decay modes of the
$\tau $--lepton and the corresponding $e^+e^- $ data related through
$CVC $. We improved the hadronic matrix element in comparison to
previous work by implementing low--lying resonances like $\rho,
\rho', \rho'', a_1 $ and $\omega $ mesons in the different
channels into the chiral structure. In particular we respected the
dominant role of the $a_1 $ meson as indicated by the data. We fixed
our parameters from $e^+e^- \to 2 \pi^+ 2 \pi^- $ and $\tau \to
\nu_{\tau} \omega \pi $ and obtained predictions for the other
four--pion decay modes of the $\tau $--lepton.

We have studied in detail two-- three-- and four--pion mass
distributions based our dynamical model
and compared our predictions with available data.
We find good agreement between the TAUOLA Monte
Carlo with our matrix elements and the experimental data. Fine tuning
of the parameters is left to the experimentalists and requires more
accurate data.

\section*{Acknowledgement}
We thank J. K\"uhn for his interest and informative discussions. \par
One of us (M.F.) would like to acknowledge partial support
by the HCM program under EEC contract number CHRX-CT920026.

%\newpage
\section*{Appendix}
Our model is available in the form of a FORTRAN code suitable to be used
with the Monte-Carlo Tauola. The file HCURR4PI.F contains three COMPLEX
FUNCTIONs, viz.\ BWIGA1, BWIGEPS, FRHO4 and the SUBROUTINEs CURINF,
CURINI and CURR.
The latter is the main
subroutine, which calls the other functions and routines
and which replaces the original subroutine
CURR of Tauola 2.4. Therefore the user has to delete the subroutine CURR
in Tauola 2.4 and link our HCURR4PI instead.
The subroutine CURINF prints out some general information about the new
routine CURR.
The subroutine CURINI initializes the COMMON block /TAU4PI/ with the
values we obtained in our fit.
The meaning
of the individual parameters is explained in Tab.~\ref{tabcommon}.
Alternatively, the user may choose to perform a new fit of
these parameters to experimental data.
\begin{table}[h] %%%%%%%%%%%%%%%%%%%%%%%%%%%%%%%%%%%%%%%
\caption{Input parameters in the common block TAU4PI}
\label{tabcommon}
$$
\begin{array}{ccc}
\hline
\qquad \mbox{Parameter} \qquad & \mbox{Meaning} & \mbox{Our fit}
\\
\hline
\mbox{GOMEGA} & g_\omega                      & 1.4 \\
\mbox{GAMMA1} & \gamma_1                      & 0.38 \\
\mbox{GAMMA2} & \gamma_2                      & 0.38 \\
\mbox{ROM1}   & m_{\rho'}\,[\mbox{GeV}]       & 1.35 \\
\mbox{ROG1}   & \Gamma_{\rho'}\,[\mbox{GeV}]  & 0.3\\
\mbox{BETA1}  & \beta_1                       & 0.08\\
\mbox{ROM2}   & m_{\rho''}\,[\mbox{GeV}]      & 1.70\\
\mbox{ROG2}   & \Gamma_{\rho''}\,[\mbox{GeV}] & 0.235\\
\mbox{BETA2}  & \beta_2                       & -0.0075\\
\hline
\end{array}
$$
\end{table}%%%%%%%%%%%%%%%%%%%%%%%%%%%%%%%%%%%%%%%%%

To obtain a copy of the file HCURR4PI.F, please send an e-mail to
M.\ F.:\newline
{\tt finkemeier@vaxlnf.lnf.infn.it}
or {\tt mf@ttpux2.physik.uni-karlsruhe.de}


\newpage
\begin{thebibliography}{99}
%
\bibitem{helt} B.K. Heltsley, Nucl. Phys. B (Proc. Suppl.) 40 (1995) 413
\bibitem{tsai} Y. S. Tsai, Phys. Rev. D 4 (1971) 2821.
\bibitem{gil}  F. J. Gilman and D. H. Miller, Phys. Rev. D 17 (1978)
               1846; F. J. Gilman and S. H. Rhie, Phys. Rev. D 31
               (1985) 1066.
\bibitem{eidel} S.I. Eidelman and V.N. Ivanchenko, Phys. Lett. B257
               (1991) 437; Nucl. Phys. B (Proc. Suppl.) 40 (1995) 131
\bibitem{arg}  ARGUS Collab., H. Albrecht et al., Phys. Lett. B 185
               (1987) 223; B 260 (1991) 259.
\bibitem{cleo} CLEO--Collab., M. Procario et al., Phys. Rev. Lett.
               70 (1993) 207; Douglas F. Cowen, CALT--68--1934.
\bibitem{aleph} P. Bourdon, Nucl. Phys. B (Proc. Suppl.) 40 (1995) 203
\bibitem{rog}  R. Decker, E. Mirkes, R. Sauer and Z. Was,  Z. Phys.
               C 58 (1993) 445.
\bibitem{tauo} S. Jadach, Z. Was, R. Decker and J. H. K\"uhn,
               Comput. Phys. Commun. 76 (1993) 361.
\bibitem{wess} R. Fischer, J. Wess and F. Wagner, Z. Phys. C 3 (1980)
               313.
\bibitem{pham} T. N. Pham, C. Roiesnel and Trang N. Truong, Phys.
               Lett. B 80 (1978) 119.
\bibitem{russen} M. Terent'ev, Sov. Phys. Uss., {\bf 17} (1974) 20;\\
               E.L. Bratkovskaya, E.A. Kuraev, Z.K. Silagadze,
               O.V. Teryaev, ``Spin-Momentun Correlation (Handedness
               in the Process of Four Pions Productions in the
               Electron-Positron Collisions'', JINR-E4-94-234 (1994);\\
	       S.I. Eidelman, Z.K. Siladadze, E.A. Kuraev,
               ``$\rho \to 4\pi$ decay'', BUDKERINP-94-87 (1994)
\bibitem{pais} A. Pais, Ann. of Phys. 9 (1960) 548.
\bibitem{pdg}  Review of Particle Properties, Phys. Rev. D 45 (1992)
               vol. 11
\bibitem{nowo} V. Sidorov , in: Proceedings of the 1979 International
               Symposium on Lepton and Photon Interactions at High
               Energies, Fermilab, editet by T.B.W. Kirk and H.D.I.
               Abarnel (Fermilab, Batavia, Illinois, 1980)
\bibitem{frasci} C. Bacci {\em et al.}, Phys. Lett. 95B (1980) 139;
                C. Bacci {\em et al.}, Nucl. Phys. B 152 (1981) 31.
\bibitem{orsay} DM2 Collaboration (D. Bisello, et al.),
                ``PWA of the $e^+ e^- \to \pi^+ \pi^- \pi^+ \pi^-$
                reaction in the $\rho'(1600)$ mass range'',
                Presented at Hadron 91, College Park, MD, Aug 12-16,
                1991.
                In: ``College Park 1991, Proceedings, Hadron '91''
                84-89 and Orsay Lin. Accel. Lab. - L.A.L. 91-64
                (91/11,rec.Mar.92) 8 p.
\bibitem{clegg} A. Donnachie and A. B. Clegg, ``A Comparative Analysis
                of $\tau$-lepton Decay and Eletron-Positron
                Annihilation'', M/C--TH 94/08
                (May 1994).
\bibitem{tau8} Review of Particle Properties, Phys. Lett. B239
               (1990) 1
\bibitem{sm}   S. L. Glashow, Nucl. Phys. 22 (1961) 579; \newline
               S. Weinberg, Phys. Rev. Lett. 19 (1967) 1264; \newline
               A. Salam, in ``Elementary particle physics:
               relativistic groups and analyticity'', ed. N.
               Svartholm (Almqvist and Wiksell, Stockholm, 1968)
\bibitem{isg}  N. Isgur, Phys. Rev. D 39 (1989) 1357.
\bibitem{kuhn} J. H. K\"uhn and A. Santamaria, Z. Phys. C 48 (1990)
               445.
\bibitem{rog2} R. Decker, Z. Phys. C 36 (1987) 487.
\end{thebibliography}
\end{document}